\documentclass{aastex631}

\usepackage{graphicx}
\usepackage{amsmath}
\usepackage{amssymb}
\usepackage{bm}
\usepackage{natbib}
\usepackage[normalem]{ulem}

\usepackage{xcolor}
\usepackage{outlines}
\definecolor{darkgreen}{RGB}{0,127,0}

\newcommand{\grad}{\boldsymbol{\nabla}}
\newcommand{\cross}{\boldsymbol{\times}}

\newcommand{\del}[2]{\ensuremath{\frac{\partial #1}{\partial #2}}}

\providecommand{\e}[1]{\ensuremath{\times 10^{#1}}}

\received{\today}
\revised{\today}
\accepted{\today}
\submitjournal{ApJ}

\shorttitle{STITCH: A Model for Coronal Energy Buildup}

\shortauthors{Dahlin et al.}

\begin{document}

\title{STITCH: A Subgrid-Scale Model for Energy Buildup in the Solar Corona}

\correspondingauthor{J.\ T.\ Dahlin}
\email{joel.t.dahlin@nasa.gov}

\author{J.\ T.\ Dahlin}


\author{C.\ R.\ DeVore}

\author{S.\ K.\ Antiochos}
\affil{Heliophysics Science Division, NASA Goddard Space Flight Center, Greenbelt, Maryland 20771}



\begin{abstract}

The solar corona routinely exhibits explosive activity, in particular coronal mass ejections and their accompanying eruptive flares, that have global-scale consequences. 
These events and their smaller counterparts, coronal jets, originate in narrow, sinuous filament channels. 
The key processes that form and evolve the channels operate on still smaller spatial scales and much longer time scales, culminating in a vast separation of characteristic lengths and times that govern these explosive phenomena. 
In this article, we describe implementation and tests of an efficient subgrid-scale model for generating eruptive structures in magnetohydrodynamics (MHD) coronal simulations. 
STITCH -- STatistical InjecTion of Condensed Helicity -- is a physics-based, reduced representation of helicity condensation: a process wherein small-scale vortical surface convection forms ubiquitous current sheets, and pervasive reconnection across the sheets mediates an inverse cascade of magnetic helicity and free energy, thereby forming the filament channels. 
STITCH abstracts these complex processes into a single new term, in the MHD Ohm's law and induction equation, which directly injects tangential magnetic flux into the low corona. 
We show that this approach is in very good agreement with a full helicity-condensation calculation that treats all of the dynamics explicitly, while enabling substantial reductions in temporal duration especially, but also in spatial resolution. 
In addition, we illustrate the flexibility of STITCH at forming localized filament channels and at energizing complex surface flux distributions that have sinuous boundaries. 
STITCH is simple to implement and computationally efficient, making it a powerful new technique for event-based, data-driven modeling of solar eruptions. 

\end{abstract}



\section{Introduction} \label{sec:intro}

Solar eruptions are the most energetic explosive events in the solar system. The largest of these events, 
known as coronal mass ejections (CMEs), release vast quantities of solar plasma and magnetic fields into the solar wind, generating shocks that disturb the entire heliosphere, driving hazardous energetic particles and magnetospheric storms at Earth.
A major goal of solar modeling efforts is operational prediction of the events that present the greatest threats to human assets and technology. A critical obstacle to achieving this goal, however, is 
the paucity of quantitative measurements of the coronal magnetic field. Further complicating matters are the remarkable
diversity of eruptions in size, from CMEs that fill the heliosphere down to coronal jets that span only a few Mm \citep{Raouafi16}; 
the morphology of the erupting material in coronagraphs including both so-called stealth CMEs \citep{Ma10} and halo CMEs \citep{Howard82}; and the 
accompanying signatures in the corona, in particular the flare ribbons, which can be as simple as the classic, linear two-ribbon structure or as complex as circular \citep{Masson09} or multi-ribbon \citep{Wang14} configurations. 
\par

Despite such diversity in eruptive manifestations, the energy source is generally agreed (at least for the largest events) to be \emph{filament channels}, highly stressed magnetic structures that form above reversals in the photospheric
line-of-sight field known as polarity inversion lines or \emph{PILs} \citep{Gaizauskas98,Martin98}. 
Filament channels are
force-balanced structures consisting of a highly sheared magnetic field (that is, approximately parallel to the PIL) restrained
by overlying unsheared field that links the opposite polarity fluxes to either side of the PIL. An eruption occurs
when this force balance is disrupted by some trigger, either an ideal instability \citep{Fan01,Linker03b,Kliem06} or magnetic reconnection \citep{Antiochos98,Antiochos99,Moore01}, which initiates 
ejection of the built-up shear field. Understanding
how the eruption will occur and the hazard it presents reduces to understanding how the generic filament-channel formation
and eruption processes occur in the complex and diverse magnetic-field configurations that house the PIL.
\par

Several models have been proposed for filament-channel formation. The simplest example, perhaps, is large-scale
shear flows or sunspot/active-region rotation \citep[e.g.,][]{Sun12,Cheung12}. However, such large-scale coherent
motion is relatively uncommon and cannot explain all such examples of filament-channel formation. For the quiet Sun especially, it has often been suggested that a combination of differential rotation and flux cancellation generate coronal flux ropes that constitute filaments and their channels \citep{Ballegooijen89,Ballegooijen04}. However, the chirality of these structures, if arising from differential rotation alone, has been shown to be inconsistent \citep[][and references therein]{Mackay14,Mackay18} with the global helicity patterns observed at all scales within eruptive coronal structures \citep[e.g.][]{Pevtsov94,Ouyang17}.
\par

The recently developed \emph{helicity condensation} model \citep{Antiochos13} posits injection of helicity/energy
into the corona via the small-scale flows associated with photospheric convective motions. In a sufficiently
``turbulent'' system supporting widespread reconnection across ubiquitous small-scale current sheets, this structure and associated energy will be transferred to increasingly larger scales via the well-known
inverse cascade of helicity \citep{Berger84a,Berger84b,Finn85,Biskamp93}. In the context of the Sun, this cascade causes small-scale twist injected
throughout an area enclosed by a PIL to eventually ``condense'' at that PIL in the form of a large-scale shear.  This theory has been rigorously explored through MHD simulations that demonstrate the formation of filament-channel structures consistent with those observed on the Sun \citep{Zhao15,Knizhnik15,Knizhnik17a,Knizhnik17b,Knizhnik18} 
and evolution of the structure through the generation of a fast CME with an eruptive flare \citep{Dahlin19}. Importantly, the filament channels generated by the helicity condensation process are consistent with the observed chirality patterns on the Sun \citep{Mackay14,Mackay18}. 
\par

All previous full-MHD studies of helicity condensation have used close-packed, boundary-driven cyclonic flows to represent helicity injection via convective granule/supergranule motions. Such calculations are computationally expensive, as they must resolve all of the cyclonic flows as well as the small-scale reconnection
dynamics that drive the helicity cascade. Additionally, the flows cannot overlap with the PILs lest they trigger excessive small-scale diffusion across structures generated at the grid scale, a challenging constraint for driving the highly complex magnetic configurations typical of eruptive events. An important implication of the helicity condensation model, however, is its relative insensitivity to the particular mechanism or spatial scale of the helicity injection \citep{Knizhnik17a,Knizhnik19}. So long as there is a net helicity injection with sufficiently complex turbulent dynamics, the shear will be transferred to the PILs. Hence, while resolving the small-scale reconnection is critically important for understanding nanoflare-type coronal heating, ultimately the large-scale energy buildup is insensitive to the underlying details of the helicity injection
and cascade processes. This motivates the search for a method that captures the essential large-scale features of the energy buildup without explicitly calculating all of the small-scale structure and dynamics. 
Such a method, if sufficiently flexible in its implementation, should be agnostic with respect to the particular mechanism for helicity injection. Hence, it could be used more generally, even in cases where turbulent convective motions are not expected to be the dominant source of energy and helicity.
\par

A statistically averaged approximation to the full helicity-condensation process was employed by \citet{Mackay14,Mackay18} to 
investigate the formation of filament channels across the full Sun and over multiple solar rotations. Those studies used a magnetofrictional model for the corona that accurately reproduces the large-scale and long-term formation of filament channels, but which cannot self-consistently capture explosive eruptive dynamics. In this article, we extend and test that statistical approach for general use in full MHD simulations. The resulting model, which we call STITCH (STatistical InjecTion of Condensed Helicity), has three critical advantages over the MHD treatment of the full helicity-condensation process with all its attendant small-scale dynamics and structures. First, it is numerically efficient in both memory and computations, requiring substantially lower spatial resolution and temporal cadence. Second, it is simple and ease to use. Third, it is highly flexible, and it is readily applicable to the complex magnetic-field distributions observed on the Sun. 
\par

We discuss the STITCH model and its numerical implementation in \S\ref{sec:method}. Our magnetohydrodynamics code, ARMS, and initial configuration are described in \S\ref{sec:simulation}. We compare STITCH to the comprehensive helicity-condensation model in \S\ref{sec:comparison}. Thereafter, we show examples of the flexibility of STITCH in localizing the filament channel (\S\ref{sec:localization}) and in being used to energize more complex and realistic magnetic configurations (\S\ref{sec:complex}). 
To conclude, we summarize our results and discuss the implications of using STITCH in simulations of coronal energy buildup in \S\ref{sec:discussion}.
\par

\section{STITCH} \label{sec:method}

In his exposition of the helicity condensation model, \citet{Antiochos13} described how an ensemble of cyclonic motions at the photosphere builds up magnetic shear in the corona. Current sheets form at the interfaces between pairs of neighboring cyclonic cells that have the same sense (clockwise or counter-clockwise) of rotation. Magnetic reconnection across the sheets depletes the twist component of magnetic flux between the cells, leaving behind the twist flux that envelopes both cells jointly. This process repeats in an inverse cascade that transports the twist flux to ever-larger spatial scales. The cascade halts at the boundary of the flux system -- defined by a reversal of either the sign of the vertical magnetic field or the sense of rotation of the cyclonic cells -- because the twist flux has the same direction on both sides of such a boundary and, hence, cannot reconnect away. The twist flux encircling the flux system is said to ``condense'' at that boundary, and the linkage between the twist flux and the enclosed vertical flux imparts magnetic helicity to the structure. The term ``helicity condensation'' had been coined previously to describe this process when it occurs in laboratory experiments and numerical simulations \citep{Biskamp93}.
\par

Several subsequent investigations of helicity condensation as applied to the corona have been carried out \citep{Zhao15,Knizhnik15,Knizhnik17a,Knizhnik17b,Knizhnik18,Dahlin19} in which numerous small-scale cyclonic cells -- as many as 100$+$ -- were resolved on the computational grid. The needed helicity-preserving simulations were performed with the Adaptively Refined Magnetohydrodynamics Solver \citep[ARMS;][]{DeVore08}. The processes of twist-flux accumulation, current-sheet formation and reconnection, and helicity condensation that were predicted theoretically were confirmed quantitatively and in detail. However, the required well-resolved, long-duration calculations of this type are very resource-intensive.
\par

In a separate line of investigation that applied the helicity condensation concept to the full-Sun magnetic field over many weeks of evolution \citep{Mackay14,Mackay18}, a subgrid-scale representation of the model was developed and employed. The essential assumptions that were made are: (1) the cyclonic cells are small, numerous, and nominally identical, although their properties may exhibit large-scale variations; and (2) the reconnection across current sheets between the cells is so efficient that the oppositely directed twist fields can be treated as simply canceling algebraically. As derived in the Appendix to the paper by \citet[][Eq.\ A1]{Mackay14}, the resulting subgrid-scale model is expressed as an additional term in the induction equation,
\begin{align}
  \del{\mathbf{B}}{t} = \grad \cross \del{\mathbf{A}_s}{t} = - \grad \cross \grad_s \left( \zeta B_n \right).
  \label{eqn:helconterm}
\end{align}
Here $\mathbf{B}$ is the magnetic field, $\mathbf{A}$ is the vector potential, subscripts $n$ and $s$ denote components normal and parallel to the surface, respectively, and $\zeta$ is a parameter characterizing the cyclonic flows. Specifically,
\begin{align}
  \zeta = \left\langle \ell^2 \omega_l \right\rangle / 2,
  \label{eqn:zeta}
\end{align}
where $\ell$ is the radius and $\omega_l$ is the angular rotation rate of the cells, and the angle brackets denote a local average in space and time. From the form of Equation \ref{eqn:helconterm}, it is clear that the changes in $\mathbf{B}$ are strongest where the product $\zeta B_n$ varies most rapidly across the surface. In particular, this occurs at the flux-system boundaries described previously. We refer to the subgrid-scale model represented by Equation \ref{eqn:helconterm} as STatistical InjecTion of Condensed Helicity, or STITCH.
\par

The STITCH term has several notable properties. First and most fundamentally, by construction it preserves the divergence-free character of the magnetic field, 
\begin{align}
  \del{}{t}\grad \cdot \mathbf{B} = - \grad \cdot \left[ \grad \cross \grad_s \left( \zeta B_n \right) \right] = 0,
  \label{eqn:helcondivb}
\end{align}
subject only to the requirement that $\zeta B_n$ be sufficiently smooth (i.e., differentiable) everywhere. Second, it prescribes an injection of twist flux into the corona that leaves the normal magnetic field component $B_n$ unchanged,
\begin{align}
  \del{B_n}{t} = - \hat{\mathbf{n}} \cdot \grad \cross \grad_s \left( \zeta B_n \right) = - \hat{\mathbf{n}} \cdot \grad_s \cross \grad_s \left( \zeta B_n \right) \equiv 0.
  \label{eqn:helconnorm}
\end{align}
The change in the horizontal field component ${\mathbf{B}_s}$ is 
\begin{align}
  \del{\mathbf{B}_s}{t} = - \grad_n \cross \grad_s \left( \zeta B_n \right) = + \grad_s \cross \grad_n \left( \zeta B_n \right);
  \label{eqn:helcontang}
\end{align}
we note that the exchange of curl and gradient operators in Equation \ref{eqn:helcontang} is valid in both spherical and Cartesian coordinates. Third, the total rate of injection of relative helicity, $H_r$, into the atmosphere is given by the surface integral of the local rate due to the individual cyclonic cells \citep[for details, see][]{Mackay14},
\begin{align}
  \frac{dH_r}{dt} = - 2 \int\!\!\!\int \zeta B_n^2 d\sigma.
  \label{eqn:helconsurf}
\end{align}
This is required in order for the STITCH term to be consistent with helicity conservation throughout the processes of reconnection and transport of twist flux to the flux-system boundaries.
\par

STITCH is implemented numerically by computing the rate at which horizontal magnetic flux is injected into the domain according to Equation \ref{eqn:helcontang}. This flux is calculated on the cell faces oriented perpendicular to the two horizontal coordinates $s_1$ and $s_2$, where $\hat{\mathbf{s_1}} \cross \hat{\mathbf{s_2}} = \hat{\mathbf{n}}$. Integrating Equation \ref{eqn:helcontang} over differential cell areas $dA_{s_1}$ and $dA_{s_2}$, respectively, we obtain the flux change rates 
\begin{align}
  \del{{\Phi}_{s_1}}{t} &= \int\!\!\!\int dA_{s_1} \phantom{.} \hat{\mathbf{s_1}} \cdot \del{\mathbf{B}_s}{t} = + \int\!\!\!\int ds_2 \phantom{.} dn \phantom{.} \del{}{s_2} \phantom{.} \del{}{n} \left( \zeta B_n \right),\label{eqn:helcontangf1}\\
  \del{{\Phi}_{s_2}}{t} &= \int\!\!\!\int dA_{s_2} \phantom{.} \hat{\mathbf{s_2}} \cdot \del{\mathbf{B}_s}{t} = - \int\!\!\!\int ds_1 \phantom{.} dn \phantom{.} \del{}{s_1} \phantom{.} \del{}{n} \left( \zeta B_n \right).\label{eqn:helcontangf2}
\end{align}
These expressions are coordinate-system independent. We localize the angular rotation of the convection cells near the coronal base, which is positioned at height $n = n_0$, by assuming that $\zeta$ falls to zero at the top of the first layer of grid cells at height $n = n_1$. 
Integrating Equations \ref{eqn:helcontangf1} and \ref{eqn:helcontangf2} over $n$ from $n_0$ to $n_1$, they simplify to 
\begin{align}
  \del{{\Phi}_{s_1}}{t} &= - \int ds_2 \phantom{.} \del{}{s_2} \left( \zeta B_n \vert_{n_0} \right),\label{eqn:helcontangf3}\\
  \del{{\Phi}_{s_2}}{t} &= + \int ds_1 \phantom{.} \del{}{s_1} \left( \zeta B_n \vert_{n_0} \right),\label{eqn:helcontangf4}
\end{align}

where $\zeta B_n \vert_{n_0}$ denotes the value of $\zeta B_n$ evaluated at the surface, $n = n_0$. These expressions integrate immediately to yield the flux changes 
\begin{align}
  \del{{\Phi}_{s_1}}{t} &= - \Delta_{s_2} \left( \zeta B_n \vert_{n_0} \right),\label{eqn:helcontangf5}\\
  \del{{\Phi}_{s_2}}{t} &= + \Delta_{s_1} \left( \zeta B_n \vert_{n_0} \right),\label{eqn:helcontangf6}
\end{align}
where $\Delta_{s_1}$ ($\Delta_{s_2}$) denotes a finite difference whose argument is evaluated at cell vertices along coordinate $s_1$ ($s_2$).
\par

The flux changes on the four faces of any grid cell involve pairwise differences of the argument $\left( \zeta B_n \vert_{n_0} \right)$ evaluated at the four vertices of the $(s_1,s_2)$ grid. Summing these changes algebraically, to calculate the change in the net flux leaving the cell through its four faces, yields zero; each vertex contribution appears twice in the sum with opposite signs ($+,-$).  This is simply a discrete expression of Gauss's law, Equation \ref{eqn:helcondivb}, integrated over the cell volume. 
\par

The final step in the calculation is to update the horizontal magnetic field components using the flux changes in Equations \ref{eqn:helcontangf5} and \ref{eqn:helcontangf6}, 
\begin{align}
  \del{{B}_{s_1}}{t} &= \frac{1}{A_{s_1}} \del{{\Phi}_{s_1}}{t} = - \frac{1}{A_{s_1}} \Delta_{s_2} \left( \zeta B_n \vert_{n_0} \right),\label{eqn:helcontangb1}\\
  \del{{B}_{s_2}}{t} &= \frac{1}{A_{s_2}} \del{{\Phi}_{s_2}}{t} = + \frac{1}{A_{s_2}} \Delta_{s_1} \left( \zeta B_n \vert_{n_0} \right),\label{eqn:helcontangb2}
\end{align}
where $A_{s_1}$ ($A_{s_2}$) is the area of the cell face perpendicular to the coordinate $s_1$ ($s_2$). Evaluating the $\left( \zeta B_n \vert_{n_0} \right)$ vertex values and multiplying them by the finite-difference time increment, $\Delta t$, yields the required changes to the field components, $\Delta B_{s_1}$ and $\Delta B_{s_2}$.
\par

\begin{figure}
\plotone{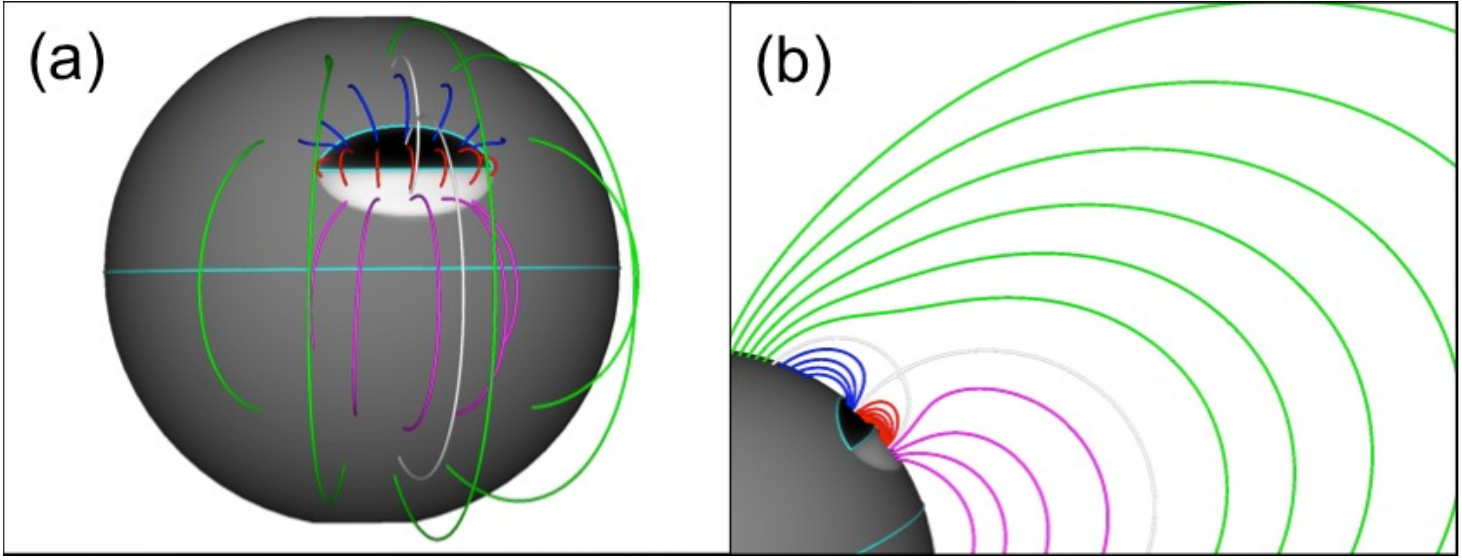}
\caption{Initial embedded-bipole configuration. Surface gray shading indicates the sign and strength of the radial magnetic field component, $B_r$; its PILs are colored cyan. Magnetic field lines are colored according to whether they close entirely within the active region (red), between the active region and the background magnetic field (blue and magenta), or entirely within the background field (green), or are separatrix lines bounding these flux systems (gray/silver).}
\label{fig:config}
\end{figure}

\section{Numerical Simulations} \label{sec:simulation}

Our numerical calculations were performed with the
Adaptively Refined Magnetohydrodynamics Solver
\citep[ARMS;][]{DeVore08}, which has been used extensively to model both
CMEs/eruptive flares \citep[see
also][]{Lynch08,Lynch09,Karpen12,Masson13,Lynch16,Lynch21} and the formation
of filament channels
\citep{Zhao15,Knizhnik15,Knizhnik17a,Knizhnik17b,Knizhnik18}. Our fiducial
calculations adopt the magnetic field configuration shown in Figure \ref{fig:config}. 
It consists of an elliptical, bipolar active region centered at 22.5$^\circ$ N latitude with peak radial field $|B_r| \approx 50$ G, embedded in a background $10$ G solar 
dipole field \citep[for full description see][]{Dahlin19}. The resulting topology is the well-known embedded
bipole with a separatrix dome and a pair of spine
lines emanating from a 3D null point \citep{Antiochos90,Lau90,Priest96}.
A region of maximally refined grid enclosed the entire separatrix dome
(gray/silver field lines arching over the arcade of red loops, Fig.\
\ref{fig:config}) and a substantial volume above it, in order to resolve the eventual small-scale reconnection dynamics. The initial atmosphere was a spherically symmetric
hydrostatic equilibrium with an inverse-$r$ temperature profile
at a base temperature $T_s = 2 \times 10^6$ K and pressure
$P_s = 4 \times 10^{-1}$ dyn cm$^{-2}$. We solved the ideal MHD equations with an adiabatic
temperature equation.
The domain 
extents were $r \in [1R_s, 30R_s]$, $\theta \in [\pi/16, 15\pi/16]$,
and $\phi \in [-\pi,+\pi]$.
\par
 
\begin{figure}
\plotone{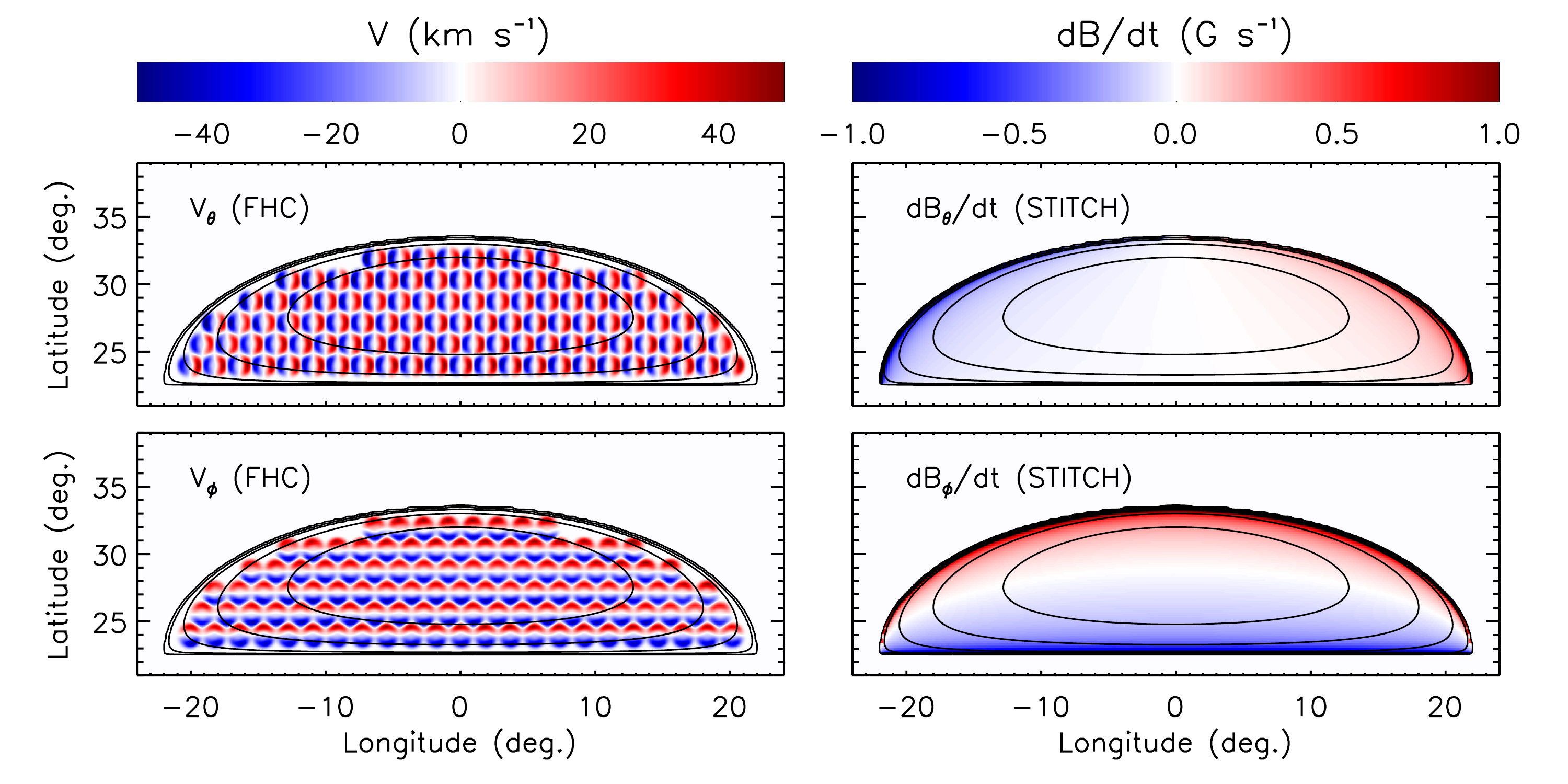}
\caption{Helicity injection for the idealized embedded-bipole configuration.
(Left) Horizontal components of velocity, $\textbf{V}_s$, are imposed on the boundary to drive the full helicity condensation (FHC) case.
(Right) Horizontal components of magnetic field, $\textbf{B}_s$, are injected just above the boundary to drive the STITCH case, here with uniform $\zeta$. Black contours indicate radial magnetic field values $B_r =$ [-50,-40,-30,-20,-10,0] G.} An animation of this figure is provided online.
\label{fig:stitch_zeta_constant}
\end{figure}

For the full helicity condensation (FHC) calculation, the system was driven by 107 tessellated, cyclonic, $B_r$-conserving
cellular flows within the black-shaded, minority-polarity region as shown in
Figure \ref{fig:stitch_zeta_constant}a. The flows were subsonic and sub-Alfv\'enic,
attaining a maximum speed $|V_\perp| \approx 50$ km s$^{-1}$ after an
initial sinusoidal ramp-up interval 1,000 s in duration. Elsewhere on the surface, the magnetic
field was line-tied at rest ($\mathbf{V_\perp} = 0$) for the
entire duration of the simulation. An extended energy-buildup phase 
occurred over about 90,000 s, which required approximately two 
continuous months of computational time on the NCCS \texttt{discover} 
supercomputer at NASA GSFC.
\par

For our first STITCH calculation described below, we assumed that $\zeta$ was uniform at $\zeta_0 = 1.4\times 10^{16}$ cm$^2$ s$^{-1}$ within the minority-polarity region, and zero everywhere outside of that region. The resulting profiles of injected horizontal magnetic field, whose maximum rate was about 1 G s$^{-1}$, are shown in Figure \ref{fig:stitch_zeta_constant}b. In this and other STITCH calculations, the magnetic field was line-tied at rest ($\mathbf{V_\perp} = 0$) over the whole surface for the entire duration of the simulation. To achieve a smooth start-up, we ramped up the tangential field injection sinusoidally over the first 1,000 s of elapsed time in the simulation. The energy-buildup phase was much shorter with STITCH, in this case being only about 6,000 s in total duration and requiring approximately four days of computational time. 
\par

We note that the stated rate of field injection and time interval suggest that a tangential field strength on the order of 6 kG should build up; however, this is not the case. The tangential field is injected just above the surface, but then it is redistributed all along the length of the filament-channel field lines by Alfv\'en waves. This reduces the accumulated maximum tangential field strength to approximately 100 G, roughly twice the peak radial field strength inside the minority polarity. Expressed another way, we find that the tangential flux injected by STITCH over the simulation's duration is roughly equal to the normal flux within the minority-polarity region. 
\par

\section{Comparing STITCH to Full Helicity Condensation}\label{sec:comparison}

\begin{figure}
\plotone{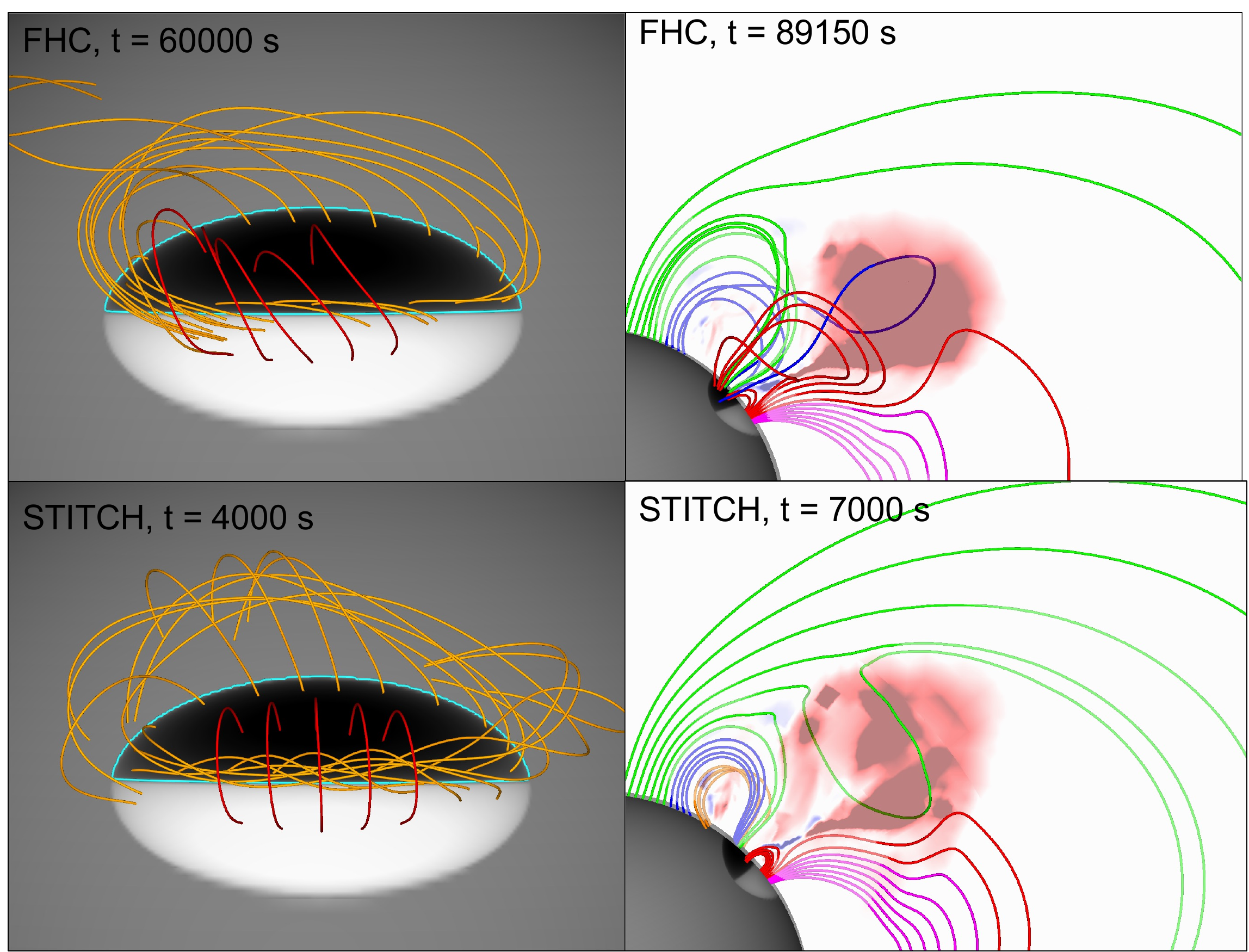}
\caption{Filament-channel formation (left) and eruption (right) for full helicity condensation (FHC; top) and STITCH (bottom) with uniform $\zeta$. Magnetic field lines are colored according to the scheme used in Figure \ref{fig:config}, with orange lines added to show the filament channel field. Color shading in the plane of the sky (right) shows radial velocity $V_r$.
}
\label{fig:hceruption}
\end{figure}

Figure~\ref{fig:hceruption} shows snapshots of the filament-channel formation and eruption phases for the FHC and STITCH cases. Both generate low-lying, highly sheared orange field lines (the longest of which stretch more than halfway along the PIL) that are restrained by overlying, weakly sheared red field lines (Fig.\ \ref{fig:hceruption}a,c). These snapshots are taken about two-thirds of the way through the filament-channel formation phase for the two cases. Eventually, the entire filament channel erupts as a quasi-circular filament with maximum speed $V > 1000$ km s$^{-1}$, indicated by dark red shading in the plane of the sky (Fig.\ \ref{fig:hceruption}b,d). The kinetic energy is near its maximum value at the times shown for both cases. 
\par

One significant point of contrast between the two cases is the shear imparted to the red loops overlying the PIL of the active region under FHC; under STITCH, the same loops are essentially shear-free. 
The helicity-injecting flows in the former fill the minority-polarity portion of the active region, inducing twist everywhere that subsequently condenses at the PIL, but with a finite time delay and with a spatial distribution that is smoothed across the region of flow. 
In contrast, the STITCH model with uniform $\zeta$ instantaneously injects tangential field adjacent to the PIL, across a zone whose characteristic width $w = \vert B_r \vert / \vert \grad_s B_r \vert$ is determined solely by the magnetic-field profile. 
The constant stirring of the field within the filament channel by the helicity-condensation flows means that the effective minimum width of the FHC channel will be the characteristic diameter of the vortical cells, $d = 2\ell$ (recall Eq.\ \ref{eqn:zeta}). 
Due to the limited resolution available for use in our simulations, these two measures of the width are comparable, $d \approx w$, as can be seen directly in Figure \ref{fig:stitch_zeta_constant}. 
Therefore, the resulting filament channels exhibit similar widths of their strongly sheared core fields, as can be seen in Figure \ref{fig:hceruption}a,c. 
\par

\begin{figure}
\plotone{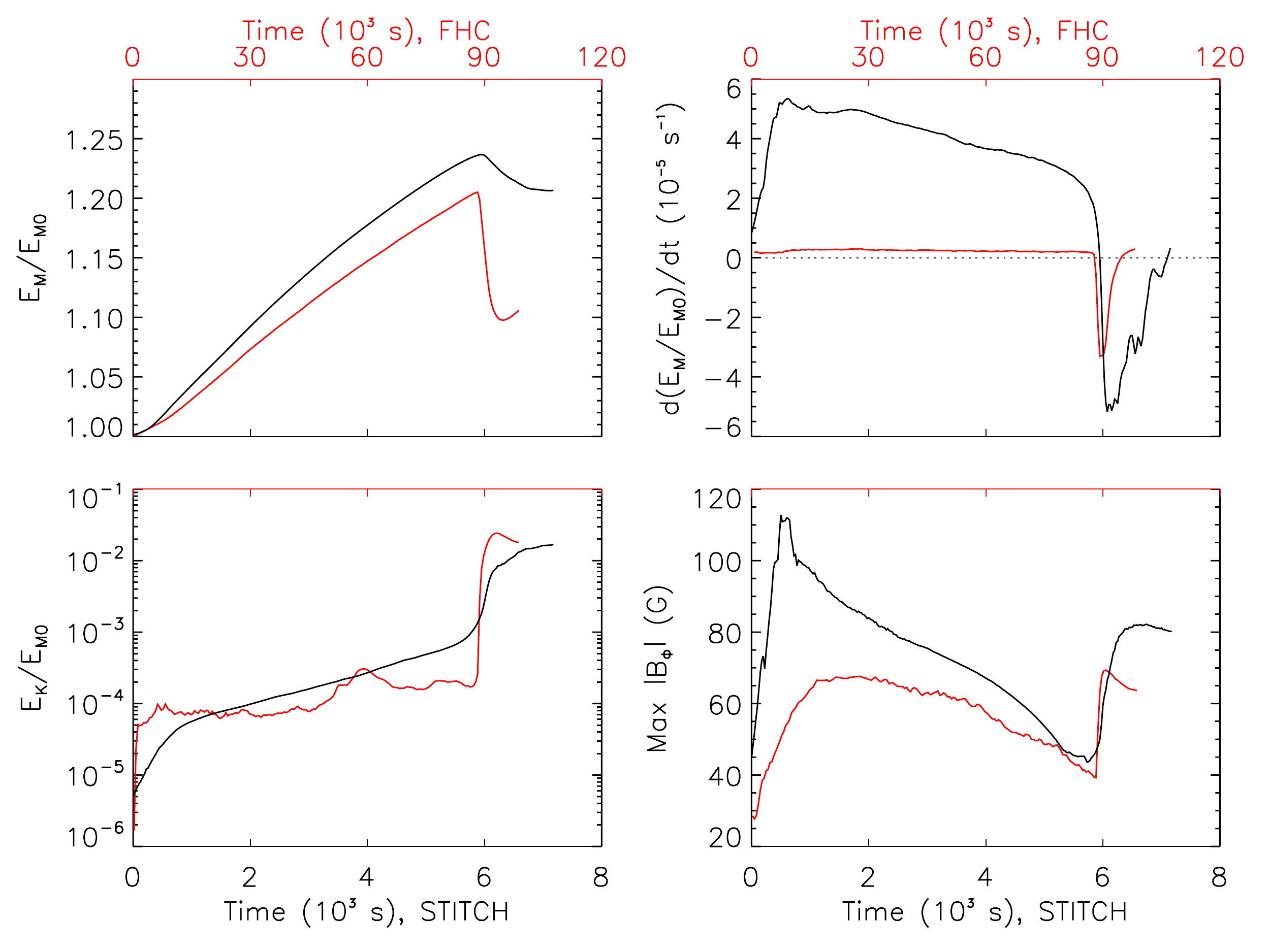}
\caption{Global diagnostics for STITCH (black curves and time-axis labels) with uniform $\zeta$ and FHC (red curves and time-axis labels): (a) $E_M/E_{M0}$; (b) $d\left(E_M/E_{M0}\right)/dt $; (c) $E_k/E_{M0}$; and (d) $\vert B_\phi \vert_{\rm max}$. 
}
\label{fig:hcenergetics}
\end{figure}

Figure~\ref{fig:hcenergetics} shows the evolution of the volume-integrated magnetic and kinetic energies, along with the maximum value of $\vert B_\phi \vert$, a proxy for the strength of the shear magnetic field.  

In order to compare the macroscopic evolution most conveniently, the respective time axes are set to align the approximate eruption times in the two cases, $t \approx$ 90,000 s (FHC) and $t \approx$ 6,000 s (STITCH).

The reduction in the nominal driving time is matched by a corresponding decrease by a factor of 15 in the computational cost of the STITCH run.
Qualitatively, all of the trends in the displayed global variables are shared between FHC and STITCH. Quantitatively, on the other hand, there are both close similarities and significant differences, as we now discuss in detail.
\par

The accumulation of magnetic energy from its initial value, $E_M \ge E_{M0}$, and its later reduction in conjunction with eruption, are shown in Figure \ref{fig:hcenergetics}a. 
We find that the accumulated magnetic free energy is some 20\% higher in the STITCH case than in the FHC case, $E_M/E_{M0} \approx$ 1.24 and 1.20, respectively. 
This STITCH excess is expected, based on its concentrating all of its shear adjacent to the PIL, which we discussed above. 
The more broadly distributed shear in the FHC case distorts the overlying coronal null point into a current patch, facilitating the onset of breakout reconnection and removal of the high-lying magnetic tethers, earlier in the energy-injection process than is true for the STITCH case. 
We emphasize that the elapsed time is so much longer for FHC only because its imposed vortical flows are so slow. 
Were we to inject magnetic free energy in the STITCH case at a comparably slow rate, we would expect the STITCH energy-buildup phase to require more elapsed time than FHC, in order to accumulate the greater energy needed to initiate the STITCH eruption. 
\par

The strong concentration of the STITCH shear adjacent to the PIL can account for two more contrasting features of its energy profile compared to the FHC case. 
First, the STITCH magnetic-energy release is only about 40\% of that for FHC, $\Delta E_M/E_{M0} \approx$ 0.04 and 0.10, respectively, culminating in post-eruption energies $E_M/E_{M0} \approx$ 1.20 and 1.10. 
The much narrower shear distribution achieved by STITCH reduces the volume of free-energy containing magnetic flux, relative to that of FHC, and throttles back the amounts of flux that is processed and energy that is released by the eruption. 
Second, the STITCH energy-release time scale is only $\Delta t \approx$ 1,000 s, a factor of four less than the $\Delta t \approx$ 4,000 s time scale for the FHC energy release. 
(Notice that the STITCH energy-release phase only appears to be much more gradual than that for the FHC phase, due to the difference in the absolute time axes for the two cases.) 
We would expect the smaller volume processed in the STITCH case to correspond to a shorter time scale, and this is exactly what we find. 
The greater concentration of the STITCH shear flux leads to a rather more explosive energy release, however, by a factor of 0.4*4.0 = 1.6, or a 60\% increase relative to the FHC energy-release rate. 
\par

For clarity, the energy storage and release rates $dE_M/dt$, normalized to the initial magnetic energy, are displayed separately in Figure \ref{fig:hcenergetics}b. 
The storage rate for STITCH is just over $15\times$ faster than that for FHC whereas, as mentioned, its release rate is only about 60\% faster. 
The energy release appears to subside about four times slower for STITCH versus FHC on this plot, due to the disparate time axes used, but in fact it subsides about four time faster, as we also noted above. 
\par

The kinetic energy in the simulation, normalized to the initial magnetic energy, is shown in Figure \ref{fig:hcenergetics}c. 
This ratio is $E_k/E_{M0} = \langle V^2 \rangle / \langle V_A^2 \rangle = \langle M_A^2 \rangle$, where $V_A$ and $M_A$ are the Alfv\'en speed and Alfv\'en Mach number, respectively. 
In both cases, the ratio during the long phase of slow driving is on the order of 10$^{-4}$, or $\langle M_A \rangle \approx 1\%$. 
The value jumps steeply during the explosive eruption phase to a few times 10$^{-2}$, or $\langle M_A \rangle \approx 15$-$20\%$. 
This rather high range highlights the pervasive Alfv\'enic outflows of the fast eruption, whose plasma occupies only a small fraction of the total volume. 
The FHC case has a larger volume of ejected flux and a higher peak kinetic-energy content, some 50\% more, than the STITCH case. 
\par

The shear-field proxy, $\vert B_\phi \vert_{\rm max}$, is shown in Figure \ref{fig:hcenergetics}d. 
This evolution, discussed previously in our work on the FHC simulation \citep{Dahlin19}, exhibits a strong initial increase as the shear flux begins to accumulate above the PIL in the low corona. 
Eventually, although the total amount of shear flux continues to rise, the peak shear strength begins to decline, as the accumulating shear magnetic pressure inflates the overlying arcade loops and expands the volume occupied by the flux. 
This steady decrease continues until eruption onset. 
Although much of the shear flux is ejected upward as core field of the CME flux rope, a lower-lying portion is left behind in the reconnected flare loops. 
These loops relax downward, and their shear-field strength increases as the occupied volume contracts. 
This generates a second, post-eruption peak in the shear proxy. 
The FHC and STITCH cases exhibit entirely similar behaviors, but due to the greater concentration of the STITCH filament channel, both of its peaks -- during the early expansion and after the eruption (110/80 G, respectively) -- are noticeably higher than those of the FHC filament channel (70/70 G). 
The local-minimum shear-field strength at eruption onset, on the other hand, is very nearly the same in the two cases (about 40 G).
\par

These results demonstrate that driving the filament channel formation process via the STITCH model produces an evolution leading to eruption that is very similar to our previous findings using full helicity condensation \citep{Dahlin19}. 
The major difference between the two is the highly compressed characteristic time scale for the STITCH energization: whereas FHC relies on slow, subsonic vortical flows that gradually build up the sheared filament channel, STITCH relies on the much faster, but subAlfv\'enic, direct injection of horizontal magnetic flux into the corona. 
A more minor, however physically very significant, difference between them is in the width of the resulting filament channel: the minimum FHC channel width is determined by the characteristic diameter $d$ of the vortical flows, whereas the STITCH channel width is determined entirely by the characteristic scale $w$ of the vertical magnetic field profile. 
The limited numerical resolution available to us for our FHC calculation produced results for a case in which $d$ is comparable to, but slightly larger than, $w$. 
In the true subgrid-scale regime where $d \ll w$, in contrast, we would expect the resulting FHC channel width to be determined by the magnetic-field length scale $w$. 
The FHC and STITCH results then should come into even better alignment than for the cases presented here, and they should show an even greater disparity in their respective time scales, further enhancing the advantages of STITCH for studies of filament channel formation. 
\par

\begin{figure}
\plotone{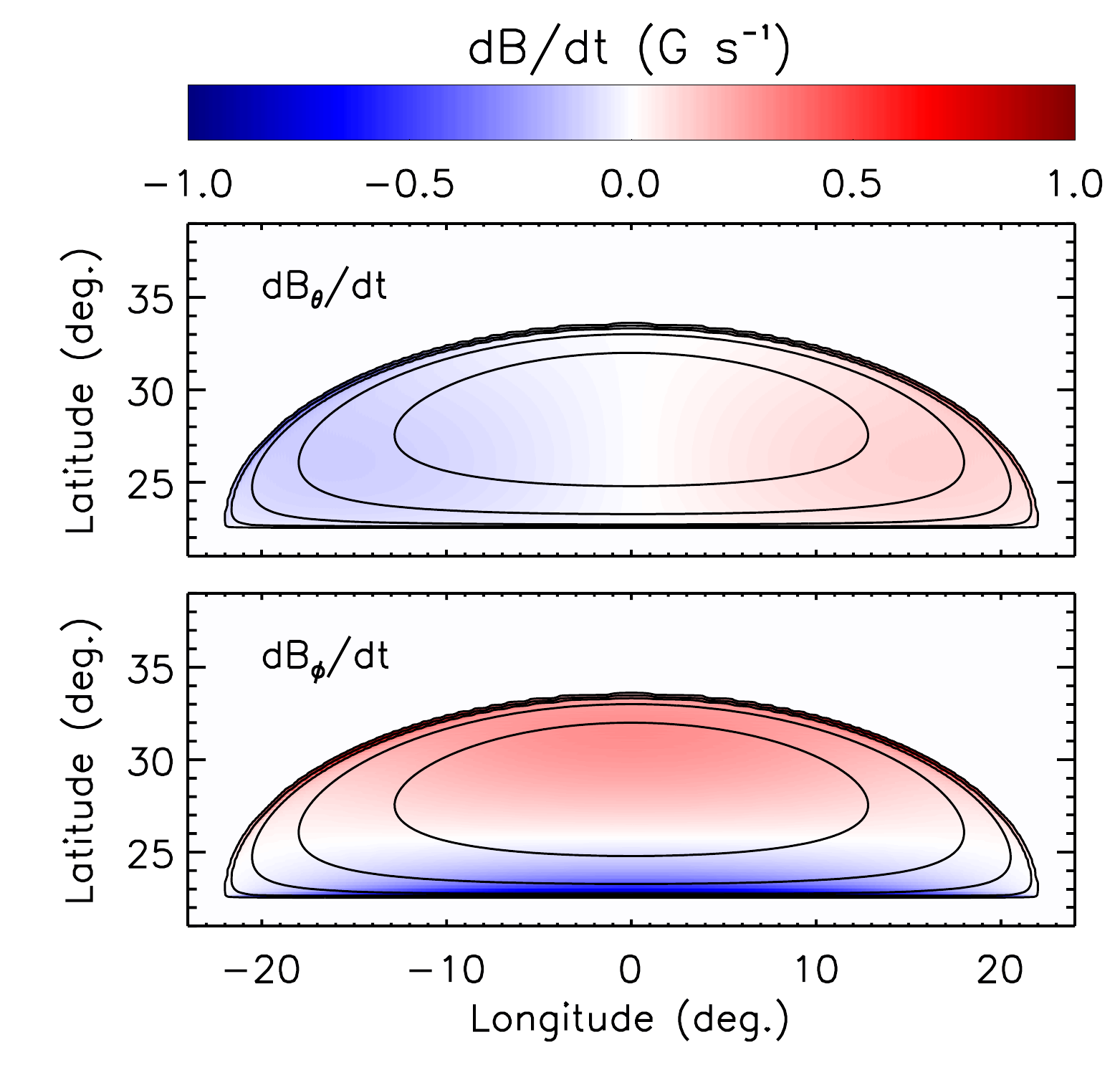}
\caption{Injection rate of tangential magnetic field components for $\zeta$ localized along the southern half of PIL (see Eq.~\ref{eqn:zeta_local}).}
\label{fig:stitch_zeta_localized}
\end{figure}

\section{Localized Helicity Injection}\label{sec:localization}

The STITCH case presented above assumed uniform $\zeta$ throughout the minority polarity region, best matching the uniform cyclonic flows in the full helicity-condensation simulation. The resulting eruptions displayed in Figure \ref{fig:hceruption} (best seen in the online animations) encompass the entire elliptical filament channel, proceeding first on the southern segment of the PIL and subsequently on the northern. This scenario is not universal, but it is observed routinely on the Sun \citep[e.g.,][]{Wang12}. The sequence implies that the energy buildup is relatively uniform all along the PIL, hence the entire sheared structure approaches its critical point and transitions to eruption nearly instantaneously. For the configuration discussed in \S\ref{sec:comparison}, for example, the gradient in $B_r$ across the PIL varies relatively little (by less than a factor of two) along the length. Therefore, the rate of STITCH flux injection likewise is quite uniform (Fig.~\ref{fig:stitch_zeta_constant}b), and the resulting eruption is a two-sided ejection of the entire channel. 
\par

\begin{figure}
\plotone{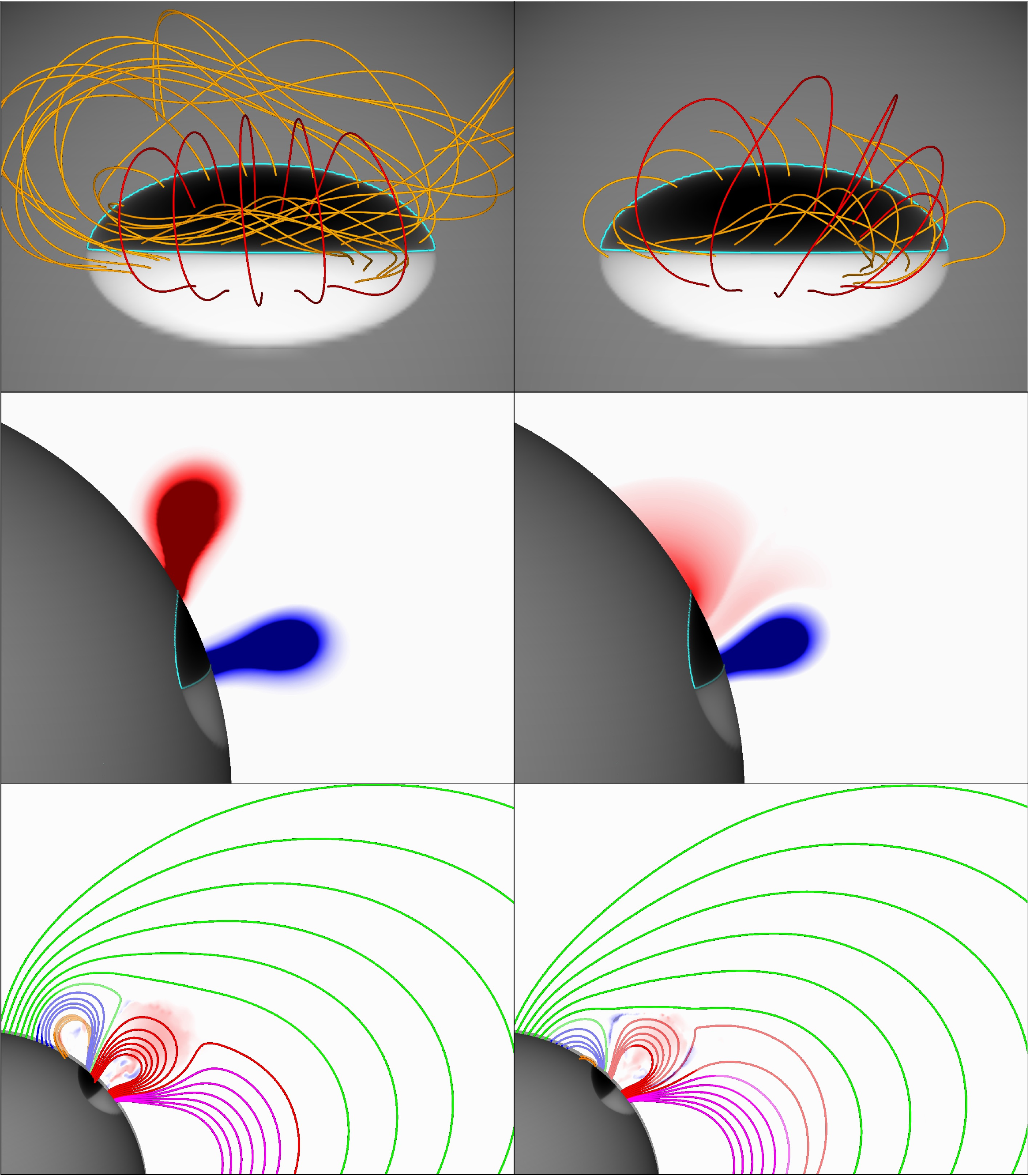}
\caption{Selected filament channel magnetic field lines plus gray-shaded radial magnetic field component (top), azimuthal magnetic field $B_\phi$ saturated at $\pm 15$ G (middle), and selected magnetic field lines plus color-shaded radial velocity component (bottom) for uniform $\zeta$ (left) and localized $\zeta$ (right) at $t =$ 6,200 s. 
An animation of this figure is provided online.
}
\label{fig:zetaeruption}
\end{figure}

More typically, however, solar eruptions occur along restricted segments of the filament channels and present one-sided ejections, even along quasi-circular PILs. 
For example, such an event is described by \citet{Mason2021}. 
The implied local concentration of energy and shear buildup could arise due to several mechanisms, e.g., variable vortical granular/supergranular and/or large-scale shear flows, disordered flux cancellation, flux emergence, etc. 
The STITCH model allows us to abstract any and all such mechanisms into our simulations by simply adopting a spatially varying $\zeta$ profile, thereby localizing the helicity and energy injection to a selected region. 
We modified our uniform-$\zeta$ simulation in this way by centering a simple cosine variation on the southern PIL, 
\begin{align}
 \zeta &= \zeta_0 \cos\left(\pi \frac{\phi-\phi_0}{\phi_w} \right) \cos\left(\pi \frac{\psi-\psi_0}{\psi_w} \right). 
 \label{eqn:zeta_local}
\end{align}
Here $\phi$ is longitude, $\psi$ is latitude, $\phi_0 = 0^\circ$, $\psi_0 = 22.5^\circ$, $\phi_w = 22.5^\circ$, and $\psi_w = 11.25^\circ$. 
The amplitude $\zeta_0$ and the restriction to the minority-polarity region are the same as in the previous simulation. 
Figure \ref{fig:stitch_zeta_localized} shows the resulting injection rates for $B_\phi$ and $B_\theta$, for comparison with those in Figure \ref{fig:stitch_zeta_constant}b. 
The principal feature is that the injection is weakened substantially along the eastern, northern, and western segments of the PIL relative to the uniform-$\zeta$ case. 
A secondary feature is that the distribution of flux injection spreads from those segments of the PIL into the interior of the minority-polarity region, due to contributions to the injection rates made by the $\zeta$ gradients. 
Based on the results of our simulation shown previously, we ramped down the tangential field injection between elapsed times $t =$ 6,000 s and $t =$ 7,000 s, reducing the injection rate during the eruption. 
We also reran the uniform-$\zeta$ case with this change in the time profile, for consistency. 
\par

Snapshots of the evolving configurations at time $t =$ 6,200 s are shown in Figure \ref{fig:zetaeruption} for the uniform-$\zeta$ (left) and localized-$\zeta$ (right) cases. 
Several significant differences are clearly evident from this comparison. 
In the uniform-$\zeta$ case, strongly sheared field lines populate the filament channel all along the PIL (Fig.~\ref{fig:zetaeruption}a); in the localized-$\zeta$ case, on the other hand, such field lines are observed only along the southern segment of the PIL, with weak shear present along the other segments by design (Fig.~\ref{fig:zetaeruption}b). 
The red arcade field lines overlying the southern filament channel are inflated to a greater extent in the localized-$\zeta$ case, due in part to less competition from the much more weakly sheared, northern segment of the PIL, and in part to the reversed shear that has been imparted to these field lines. 
This reversed shear arises from the relatively small, but finite, injection of oppositely signed $B_\phi$ flux near the center of the minority-polarity region due to the $\zeta$ gradients in the STITCH term (noted in the description of Fig.~\ref{fig:stitch_zeta_localized}b). 
\par

Side views of the azimuthal magnetic field $B_\phi$ show how localizing the STITCH injection (Fig.~\ref{fig:zetaeruption}d) greatly reduces the shear along the northern PIL segment and introduces the reversed shear in the center of the minority polarity (both shaded light red), relative to the uniform STITCH injection (Fig.~\ref{fig:zetaeruption}c). 
Obvious features of the uniform-$\zeta$ case (Fig.~\ref{fig:zetaeruption}c) are the roughly equal strengths of its northern and southern filament channels and the extents to which they bulge upward and outward as they inflate toward nearly simultaneous eruptions. 
In contrast, only the southern filament channel is clearly primed for eruption in the localized-$\zeta$ case (Fig.~\ref{fig:zetaeruption}d). 
These views also highlight a slight northward tilt of the southern filament channel field and its overlying arcade (shaded blue) in this case. 
\par

Side views of magnetic field lines in the four flux systems of the configuration further illustrate these features in the uniform-$\zeta$ (Fig.~\ref{fig:zetaeruption}e) and localized-$\zeta$ (Fig.~\ref{fig:zetaeruption}f) cases. 
The arcade of blue field lines in the north, along with its enclosed orange field lines, resides much lower in height in the latter versus the former. 
The color shading of radial velocity $V_r$ in these panels captures the outward motion of the central arcade of red field lines in both cases, and also illustrates breakout-reconnection downflows along the flanks of this arcade in the localized-$\zeta$ case (Fig.~\ref{fig:zetaeruption}f).
\par

\begin{figure}
\plotone{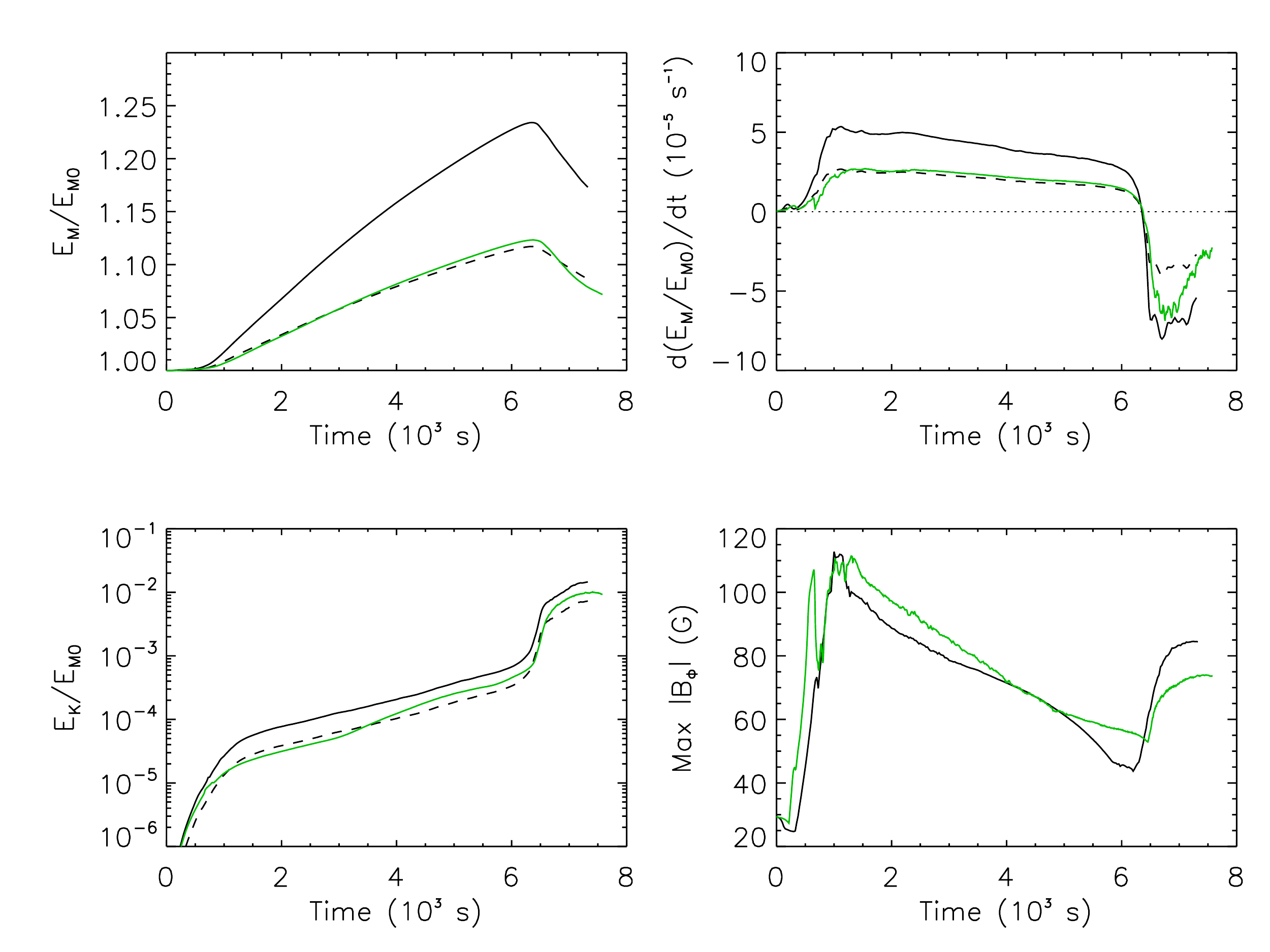}
\caption{Global diagnostics for uniform $\zeta$ (black curves) and localized $\zeta$ (green curves): (a) $E_M/E_{M0}$; (b) $d\left(E_M/E_{M0}\right)/dt $; (c) $E_k/E_{M0}$; and (d) $\vert B_\phi \vert_{\rm max}$. 
Dashed black curves in (a-c) show halved quantities for uniform $\zeta$.}
\label{fig:zetadiag}
\end{figure}

Global diagnostics for these two simulations, uniform-$\zeta$ (black curves) and localized-$\zeta$ (green curves), are compared in Figure \ref{fig:zetadiag}. 
(In these plots, unlike the previous similar Figure \ref{fig:hcenergetics}, the time axes for the two cases [both STITCH] are identical.) 
The energy Figures \ref{fig:zetadiag}a,b,c also show the uniform-$\zeta$ results at 50\% amplitude, drawn with dashed curves. 
Coincidentally, these reduced curves track very well with those for the localized-$\zeta$ case, in the buildup to eruption for the magnetic energy (Figs.~\ref{fig:zetadiag}a,b) and beyond into the eruption phase for the kinetic energy (Fig.~\ref{fig:zetadiag}c). 
This agreement indicates that, during the buildup, the northern and southern segments of the PIL store roughly equal amounts of energy for uniform $\zeta$. 
The magnetic-energy release rate during the eruption is essentially the same in the two cases, although the release subsides somewhat faster for localized $\zeta$ (Fig.~\ref{fig:zetadiag}b). 
The kinetic energy for that case peaks slightly above 50\% of the uniform-$\zeta$ value (Fig.~\ref{fig:zetadiag}c). 
Peak strengths of the shear field attained early in the energization phase are essentially identical for the two cases (Fig.~\ref{fig:zetadiag}d). 
Early in the evolution for localized $\zeta$, a transient expansion of the field overlying the channel briefly reduces the shear strength, producing a double peak. 
Late in the evolution, the shear field does not drop quite as low prior to eruption, nor does it increase quite as high after eruption, compared to the uniform-$\zeta$ case. 
These two features indicate that the filament channel and its overlying arcade neither expand quite as freely, nor contract quite as much, when the shear is localized to the southern PIL. 
\par

\section{Energizing Complex Flux Distributions}\label{sec:complex}

\begin{figure}
\plotone{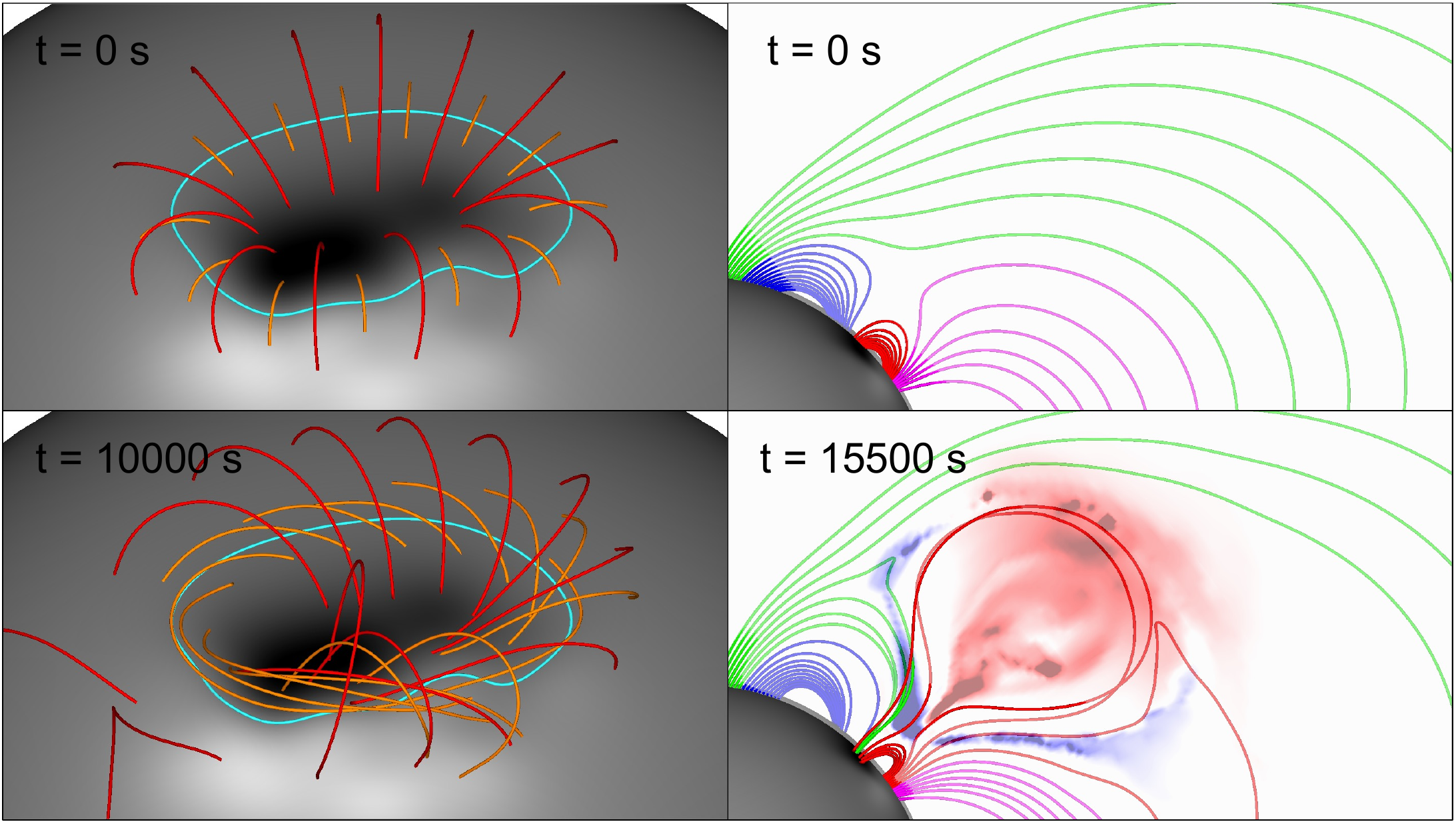}
\caption{
Initial configuration (top) of a more complex flux distribution with a corrugated PIL and later snapshot (bottom) during formation (left) and eruption (right) of its filament channel. 
Magnetic field lines are color-coded and radial velocity is color-shaded as in Figure \ref{fig:hceruption}. 
}
\label{fig:fc_complex_bipole}
\end{figure}

The idealized flux distribution used in the preceding examples has sufficed to show some of the capabilities of the STITCH model. 
Actual solar flux distributions can be much more complex, even fragmented, posing far greater challenges to data-driven modeling of such regions and requiring far greater computational resources. 
To explore the potential usefulness of STITCH for investigations of this kind, we created a synthetic active region with a corrugated PIL by superposing 30 subsurface magnetic dipoles, all centered nominally at latitude $\psi_0 = 22.5^\circ$ and longitude $\phi_0 = 0^\circ$. 
We randomized their displacements away from this center, $\vert \psi-\psi_0 \vert \le 7.5^\circ$ and $\vert \phi \vert \le 22.5^\circ$, and their horizontal orientations away from due north, within $\pm 45^\circ$. 
Each dipole was placed at a depth of 80 Mm below the surface and was given a tangential field strength of 6.7 G at the surface. 
Figure \ref{fig:fc_complex_bipole}a,b shows views of the resulting initial magnetic field with its corrugated PIL. 
\par

\begin{figure}
\plotone{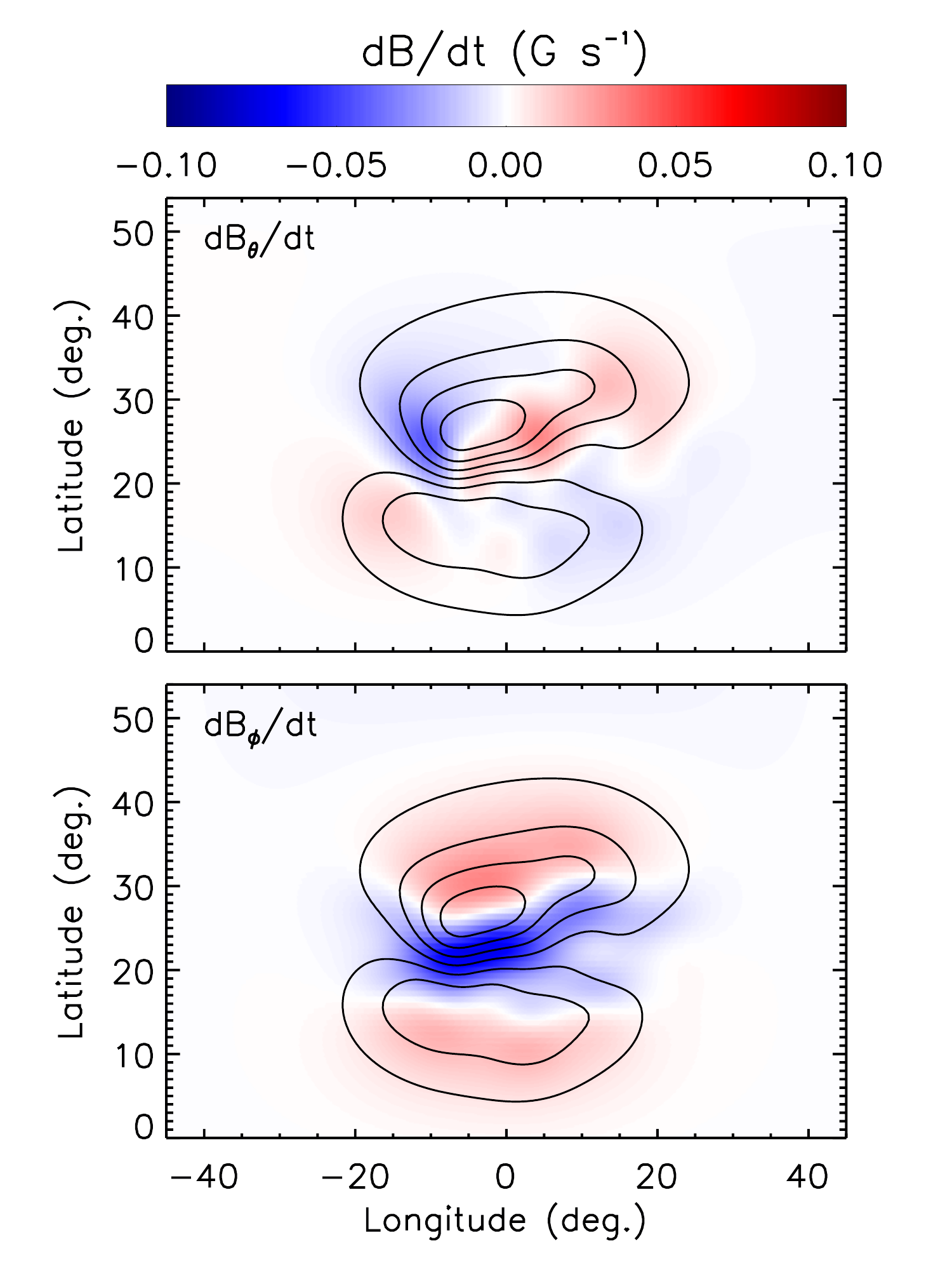}
\caption{
Tangential magnetic field injection for a more complex surface flux distribution with a corrugated PIL. 
Black contours show values of $B_r$ = [$-$30, $-$20, $-$10, 0, $+$10, $+$20] G. 
}
\label{fig:stitch_zeta_complex}
\end{figure}

The active region was energized using the STITCH profiles of tangential magnetic field injection shown in Figure \ref{fig:stitch_zeta_complex}. 
We adopted $\zeta$ as expressed in Equation (\ref{eqn:zeta_local}), but here we used $\phi_w = 45^\circ$ and $\psi_w = 27^\circ$. 
As can be seen in the figure, STITCH flux injection was applied in both polarities of the active region, as well as in the surrounding background magnetic field. 
In this case, we used a smaller value of $\zeta_0 = 5\times10^{15}$ cm$^2$ s$^{-1}$, and the gradients in the radial magnetic field $B_r$ were gentler due to our constructing the active region from a set of well-submerged dipoles. 
Consequently, the resulting values of $d\mathbf{B}_s/dt$ were about an order of magnitude smaller than in our idealized cases (cf.\ Fig.\ \ref{fig:stitch_zeta_localized}). 
We allowed the usual ramp-up interval of 1,000 s duration in this simulation, then held the tangential-field injection rates fixed thereafter. 
\par

The elapsed time required to build a strongly sheared filament channel and induce it to erupt was about 2.5$\times$ larger for this case, i.e.\ some 15,000 s. 
Snapshots of the formation and eruption stages in the evolution are shown in Figure \ref{fig:fc_complex_bipole}c,d. 
The development is very similar qualitatively to that shown in our previous examples, with the filament channel highly localized to the PIL and generating a fast CME when it erupts. The key point here is that essentially no change in the STITCH driving procedure from the simple case was required for handling this ``realistic" PIL, whereas, driving by small-scale twists or by shear flows or flux-rope insertion would have required the tedious construction of drivers that followed the detailed geometry of the PIL. 
\par

\section{Discussion\label{sec:discussion}}

STITCH is our acronym for STatiscal InjecTion of Condensed Helicity, a subgrid-scale physical model for the formation and evolution of filament channels on the Sun. 
In this article, we have used STITCH to inject magnetic helicity and free energy into simulated coronae that evolve self-consistently according to the otherwise-familiar equations of magnetohydrodynamics. 
The only modification to standard MHD that is required is the addition of a new, mathematically simple term to the induction equation (Faraday's law). 
Including the STITCH term amounts to adjusting the electric field, i.e.\ modifying Ohm's law in the coronal plasma, at the base of the atmosphere. 
\par

The mathematical simplicity of the STITCH term obscures, but simultaneously encapsulates, a great deal of physical complexity. 
As derived originally \citep{Mackay14,Mackay18}, it represents the injection of helicity by vortical cells characterized by rotation rate $\omega_\ell$ and radius $\ell$, the formation of current sheets at the boundaries between adjacent cells, and the efficient reconnection of the induced horizontal magnetic fields across the current sheets. 
The STITCH model distills these detailed processes of helicity condensation described by \citet{Antiochos13} into a single term with one parameter, $\zeta \equiv \langle \ell^2 \omega_\ell \rangle / 2$, plus an assumption about the height $h$ over which the helicity is injected into the magnetic field. 
In practice, the simplest approach is to perform the injection into the bottom-most grid cell adjacent to the lower boundary, as has been done here and by others \citep{Mackay14,Mackay18,Lynch21}. 
\par

Implementing and using the STITCH term in MHD models are straightforward: 
\begin{enumerate}
\item Assign a value to the helicity-injection parameter $\zeta$. 
\item Evaluate the STITCH product $\zeta B_n$ at the bottom boundary, $n=n_0$, where $B_n$ is the normal component of the magnetic field and $n=x$ (Cartesian) or $n=r$ (spherical). 
\item Difference this product along the two horizontal coordinates, $(y,z)$ or $(\theta,\phi)$, setting the sign according to the rules for the curl (see Equations \ref{eqn:helcontangf5} and \ref{eqn:helcontangf6}), to compute the tangential flux-change rates. 
\item Divide the flux-change rates by the appropriate vertical cell-face areas to calculate the corresponding tangential-field change rates (see Equations \ref{eqn:helcontangb1} and \ref{eqn:helcontangb2}). 
\item Multiply these rates by the numerical time increment, $\Delta t$, finishing the calculation of the tangential-field changes. 
\end{enumerate}

\par

To our knowledge, the procedure above would be trivial to implement in any of the MHD codes presently used by the space science community. 
Moreover, the helicity injection via STITCH may be tailored spatially to occur only within one polarity of the magnetic field, or only within one or more restricted region(s) of the bottom-boundary surface, or any combination of both. 
This is accomplished by introducing suitable spatial variations into the $\zeta$ parameter adopted in step \#1, as we have done in this paper. The results for the tangential-field changes will be mathematically smooth so long as the product $\zeta B_n$ goes to zero smoothly at the boundary of the tailored region, i.e., either $\zeta$ vanishes (by construction) or $B_n$ vanishes (at a PIL). 

\par

The results presented in this paper demonstrate that STITCH has a number of key advantages in comparison to other commonly used methods for filament channel formation in 3D numerical modeling of flares and solar eruptive events \citep[e.g.,][]{Patsourakos20}. These methods include flux emergence \citep[e.g.,][]{Manchester04,Leake13}, shear flows \citep[e.g.,][]{Lynch08,Wyper17}, flux cancellation \citep[e.g.,][]{Amari10,Aulanier10}, and several procedures for direct insertion of a flux rope \citep[e.g.,][]{Titov99,Ballegooijen04,Fan05,Borovikov17,Titov21}. One problem for these methods is matching the normal flux at the boundary of the numerical system to photospheric observations. This is particularly difficult for emergence, shear flows, and cancellation, because these processes generally change the flux distribution. Another challenge is dealing with PILs of complex geometry, which can pose a severe challenge for flux rope insertion, at times, requiring the specification of multiple different flux ropes carefully fitted to match an observed filament channel \citep[e.g.,][]{Torok18}. It seems unlikely that such painstaking, iterative procedures could be used for space weather predictive models. A further problem is that after an equilibrium has been calculated, the eruption must be triggered without disturbing the original normal flux distribution. STITCH inherently takes care of all these problems. In contrast to flux rope insertion, STITCH shear injection is continuous and therefore can be readily halted at any point to examine pre-eruptive structure and trigger mechanisms near marginal stability. Note also that STITCH is so flexible and easily implemented that it can be used in combination with these other filament channel formation mechanisms. For example, one could insert some equilibrium flux rope, as desired, and then use STITCH to drive the system to erupt while still matching the photospheric boundary conditions.    

\par

Although the STITCH model is derived from the intricate physical processes inherent to the full helicity condensation model, we emphasize that its essence is to accumulate helicity at the large-scale boundaries of magnetic flux systems, in particular at PILs. 
Therefore, in general it represents the outcome of an inverse cascade of magnetic helicity, from small scales to large, due to any mechanism that injects helicity into the corona: vortical motions, shearing motions, flux emergence, flux cancellation, and so forth. 
Our results in this article demonstrate that STITCH can be used as a generic tool for coronal MHD modeling that forms filament channels and energizes magnetic-field configurations. 
Here we have used it to evolve both highly idealized and more elaborately corrugated active-region structures, and we have compared it successfully, both qualitatively and quantitatively, with a fully detailed helicity-condensation calculation. 
In other work, it has been used to perform full-Sun, sunspot-cycle scale modeling of filaments in a flux-transport model \citep{Mackay14,Mackay18} and to energize a high-latitude filament channel leading to the eruption of a ``stealthy'' CME \citep{Lynch21}. 
\par

The STITCH model possesses the following efficiencies and advantages compared to fully detailed MHD descriptions of filament-channel formation and eruption: 

\begin{itemize}
\item Straightforward implementation within existing MHD models
\item Easy application to realistic, complex PILs, as well as to localized regions along extended PILs as is likely to be required for performing data-driven modeling of observed events
\item Energizes system while leaving the normal component of the magnetic field untouched, thereby, greatly facilitating event modeling
\item Reduced computational expense (15$\times$ in our detailed comparison) because STITCH is amenable to subAlfv\'enic (flux-injection) driving rather than requiring subsonic (vortical-flow) driving, allowing faster formation and eruption of filament channels
\item Reduced computational expense (not exploited in our detailed comparison) because the small-scale vortical flows and current structures need not be resolved, allowing the use of coarser grids
\item Elimination of complex small-scale structure and dynamics, so that the large-scale topology and dynamics of the filament-channel evolution can be ascertained more readily

\end{itemize}

\noindent
In summary, STITCH demonstrates substantial utility for physics-based modeling of CMEs and, potentially, great promise as an operational tool for future space-weather prediction capabilities.
\par
\par

Our work was sponsored by NASA's H-LWS, H-SR, and H-ISFM research programs and a grant of HEC computer resources to C.R.D.\ at NASA's Center for Climate Simulation. J.T.D.\ was supported by a NASA Postdoctoral Program fellowship administered by Universities Space Research Association at NASA Goddard Space Flight Center.



\newpage

\bibliography{bibliography.bib}




\end{document}